\documentclass[%
reprint,
superscriptaddress,
amsmath,amssymb,
aps,
pra,
nofootinbib
]{revtex4-2}
\usepackage[utf8]{inputenc}
\usepackage[english]{babel}
\usepackage{blindtext}
\usepackage{graphicx}   
\graphicspath{{figs/}}   
\usepackage{dcolumn}
\usepackage{bm}
\usepackage[hidelinks]{hyperref}
\usepackage{svg}
\usepackage{float}
\usepackage{epstopdf}
\usepackage{braket}
\usepackage{upgreek}
\usepackage{amsmath}
\usepackage{siunitx}
\usepackage{changes}
\usepackage{natbib}
\usepackage[title]{appendix}

\newcommand{\secref}[1]{\hyperref[#1]{{Section~\ref{#1}}}}
\newcommand{\chapref}[1]{\hyperref[#1]{{Chapter~\ref{#1}}}}
\newcommand{\suppref}[1]{\hyperref[#1]{{App.~\ref{#1}}}}
\newcommand{\figref}[1]{\hyperref[#1]{{Fig.~\ref*{#1}}}}
\newcommand{\Figref}[1]{\hyperref[#1]{{Figure~\ref*{#1}}}}
\newcommand{\figrefadd}[2]{\hyperref[#1]{{Fig.~\ref*{#1}#2}}}
\newcommand{\Figrefadd}[2]{\hyperref[#1]{{Figure~\ref*{#1}#2}}}
\newcommand{\tabref}[1]{\hyperref[#1]{\textsl{Table~\ref*{#1}}}}
\renewcommand{\eqref}[1]{\hyperref[#1]{{Eq.~\ref*{#1}}}}

\newcommand{\revisex}[1]{{}}

\usepackage{multirow}

\usepackage{import}
\usepackage{calc}
\usepackage{xcolor}
\definecolor{myBrown}{HTML}{947139}
\definecolor{myDarkgreen}{HTML}{6E692A}
\definecolor{myGreen}{HTML}{00A76D}
\definecolor{myLightgreen}{HTML}{A6CE42}
\definecolor{myYellow}{HTML}{FFF200}
\definecolor{myRed}{HTML}{ED1C1A}
\definecolor{myOrange}{HTML}{F7931D}
\definecolor{myLightblue}{HTML}{00C0F3}
\definecolor{myPink}{HTML}{F6979F}
\definecolor{myBlue}{HTML}{0071BC}
\definecolor{myGold}{HTML}{FFCB04}
\definecolor{myPurple}{HTML}{A066AA}
\definecolor{myDarkgrey}{HTML}{6D6E70}
\definecolor{myLightgrey}{HTML}{9D9FA1}

\begin{document}

\title{Pure kinetic inductance coupling for cQED with flux qubits}

\author{Simon Geisert}
\affiliation{IQMT,~Karlsruhe~Institute~of~Technology,~76131~Karlsruhe,~Germany}

\author{Soeren Ihssen}
\thanks{First two authors contributed equally.}
\affiliation{IQMT,~Karlsruhe~Institute~of~Technology,~76131~Karlsruhe,~Germany}

\author{Patrick Winkel}
\affiliation{IQMT,~Karlsruhe~Institute~of~Technology,~76131~Karlsruhe,~Germany}
\affiliation{Departments~of~Applied~Physics~and~Physics,~Yale University,~New Haven,~CT,~USA}
\affiliation{Yale~Quantum~Institute,~Yale~University,~New~Haven,~CT,~USA}

\author{Martin Spiecker}
\affiliation{IQMT,~Karlsruhe~Institute~of~Technology,~76131~Karlsruhe,~Germany}

\author{Mathieu Fechant}
\affiliation{IQMT,~Karlsruhe~Institute~of~Technology,~76131~Karlsruhe,~Germany}

\author{Patrick Paluch}
\affiliation{IQMT,~Karlsruhe~Institute~of~Technology,~76131~Karlsruhe,~Germany}
\affiliation{PHI,~Karlsruhe~Institute~of~Technology,~76131~Karlsruhe,~Germany}

\author{Nicolas Gosling}
\affiliation{IQMT,~Karlsruhe~Institute~of~Technology,~76131~Karlsruhe,~Germany}

\author{Nicolas Zapata}
\affiliation{IQMT,~Karlsruhe~Institute~of~Technology,~76131~Karlsruhe,~Germany}

\author{Simon Günzler}
\affiliation{IQMT,~Karlsruhe~Institute~of~Technology,~76131~Karlsruhe,~Germany}

\author{Dennis Rieger}
\affiliation{IQMT,~Karlsruhe~Institute~of~Technology,~76131~Karlsruhe,~Germany}
\affiliation{PHI,~Karlsruhe~Institute~of~Technology,~76131~Karlsruhe,~Germany}

\author{Denis B\'en\^atre}
\affiliation{IQMT,~Karlsruhe~Institute~of~Technology,~76131~Karlsruhe,~Germany}

\author{Thomas Reisinger}
\affiliation{IQMT,~Karlsruhe~Institute~of~Technology,~76131~Karlsruhe,~Germany}

\author{Wolfgang Wernsdorfer}
\affiliation{IQMT,~Karlsruhe~Institute~of~Technology,~76131~Karlsruhe,~Germany}
\affiliation{PHI,~Karlsruhe~Institute~of~Technology,~76131~Karlsruhe,~Germany}

\author{Ioan M. Pop}
\email{ioan.pop@kit.edu}
\affiliation{IQMT,~Karlsruhe~Institute~of~Technology,~76131~Karlsruhe,~Germany}
\affiliation{PHI,~Karlsruhe~Institute~of~Technology,~76131~Karlsruhe,~Germany}
\affiliation{Physics~Institute~1,~Stuttgart~University,~70569~Stuttgart,~Germany}

\date{\today}
\begin{abstract}
 We demonstrate a qubit-readout architecture where the dispersive coupling is entirely mediated by a kinetic inductance. This allows us to engineer the dispersive shift of the readout resonator independent of the qubit and resonator capacitances. We validate the pure kinetic coupling concept and demonstrate various generalized flux qubit regimes from plasmon to fluxon, with dispersive shifts ranging from 60\,$\textrm{kHz}$ to 2\,$\textrm{MHz}$ at the half-flux quantum sweet spot. We achieve readout performances comparable to conventional architectures with quantum state preparation fidelities of 99.7\,\% and 92.7\,\% for the ground and excited states, respectively, and below 0.1\,\% leakage to non-computational states. 
\end{abstract}
\maketitle
\maketitle

The ability to convert model Hamiltonians into programmable physical systems is a stepping stone for quantum information processing. Circuit quantum electrodynamics (cQED) has been at the forefront of quantum hardware development over the past two decades~\cite{Wallraff__cQED__2004, Blais__Review_cQED__2021}, benefitting from the freedom to design various microelectronic circuit elements such as qubits, control, readout and coupler structures from the same basic building blocks. This has led to the development of increasingly complex quantum processors \cite{Sung__Two_Tmon_SingleChip_Device__2021, Conner__Superconducting_qubits_in_a_flip_chip_architecture__2021, Kosen__Two_Tmon_FlipChip_Device__2022, Wu__Quantum_Processor__2021, Arute__Google_Processor__2019} and facilitated the exploration of fundamental quantum effects \cite{Campagne__state_diffusion__2016, Minev__To_catch_and_reverse_a_quantum_jump_mid_flight__2019, Leger__Observation_of_quantum_many_body_effects_due_to_zero_point_fluctuations_in_superconducting_circuits__2019, Stevens__Energetics_of_gate__2022, Chakram__Multimode_photon_blockade__2022, Mehta__Down_conversion_of_a_single_photon_as_a_probe_of_many_body_localization__2023, Roch__Measurement_induced_entanglement__2015}.

Dispersive coupling between qubits and harmonic oscillators is a pivotal resource for cQED, enabling single shot readout \cite{Vijay__Single_Shot_Readout__2011, Heinsoo__Single_Shot_Readout__2018, Swiadek__Single_Shot_Readout__2023, Takmakov2021Jun}, the creation of non-classical photonic states \cite{Hofheinz__Non_classical_Photonic_States__2009, Kirchmair__Non_classical_Photonic_States__2013}, reservoir engineering for qubit state preparation \cite{Geerlings__Autonomous_state_preparation__2013, Murch__Autonomous_state_preparation__2012} and even the autonomous stabilisation of entangled states \cite{Shankar__Autonomous_stabiliziation_of_entangled_states__2013}. Conventionally, dispersive coupling is mediated via electromagnetic interaction, most commonly using the electric field and a coupling capacitor. However, in complex devices, stray capacitors inevitably introduce unwanted cross-talk, renormalize the dispersive shift and even induce undesired electromotive forces across non-linear elements in the presence of alternating magnetic fields or field gradients~\cite{Lu2023Sep}. In order to reduce the number of spurious electromagnetic modes and parasitic capacitances, several mitigation strategies are currently being developed in the community, including deep silicon vias \cite{Jeffrey__An_overview_of_through_silicon_via_technology_and_manufacturing_challenges__2015, Yost__Solid_state_qubits_integrated_with_superconducting_through_silicon_vias__2020, Alfaro__Highly_Conformal_Sputtered_Through_Silicon_Vias_With_Sharp_Superconducting_Transition__2021}, flip chip architectures~\cite{Yost__Solid_state_qubits_integrated_with_superconducting_through_silicon_vias__2020, Yu__Indium_based_Flip_chip_Interconnection_for_Superconducting_Quantum_Computing_Application__2022, Kosen__Two_Tmon_FlipChip_Device__2022, Conner__Superconducting_qubits_in_a_flip_chip_architecture__2021} and chiplets~\cite{ Smith__Scaling_Superconducting_Quantum_Computers_with_Chiplet_Architectures__2022, Field__Modular_Superconducting_Qubit_Architecture_with_a_Multi_chip_Tunable_Coupler__2024}.

Here, we present an alternative coupling approach that implements dispersive readout via pure kinetic inductance coupling between a generalized flux qubit and a harmonic oscillator and enables the complete suppression of capacitive coupling. We achieve this by designing a three-island circuit with two normal modes, i.e. qubit and resonator, coupled solely by a kinetic inductance. While the kinetic inductance can be realized with Josephson junction (JJ) arrays, we demonstrate the concept with a high kinetic inductance material, namely granular aluminum (grAl)~\cite{Gru__High_Impedance_Quantum_Circuits__2019, Rieger__Gralmonium__2022}. The circuit's symmetry effectively eliminates capacitive contributions to the qubit-readout interaction, rendering the coupling local. 

\begin{figure}[!htb]
    \resizebox{1\columnwidth}{!}{ 
        \def\svgwidth{1.0\columnwidth}
        \includegraphics[width=1.0\columnwidth]{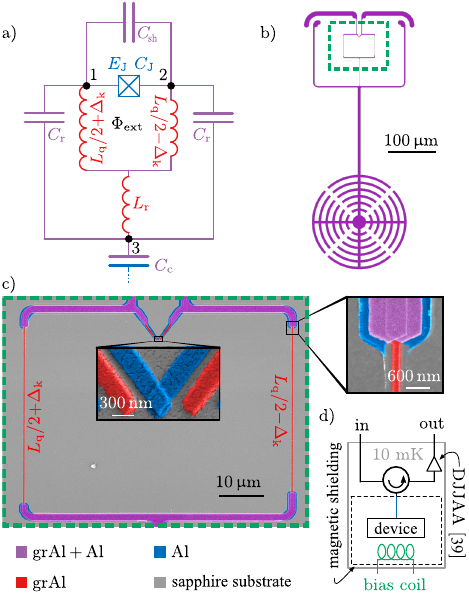}
        }
    \caption{\label{fig:GFQ} \textbf{Qubit-readout circuit schematics and implementation.} \textbf{a)} The islands of the device are labeled by the indices \textbf{1}-\textbf{3}. Islands \textbf{1} and \textbf{2} are connected by a tunnel junction with Josephson energy $E_\text{J}$ and capacitance $C_\text{J}$, shunted by a grAl inductance $L_\text{q}$ and a capacitance $C_\text{sh}$, to form a generalized flux qubit. By connecting the third island to the qubit loop via the inductance $L_{\text{r}}$, we engineer the readout mode that charges the islands \textbf{1} and \textbf{2} in-phase. The qubit-readout coupling is controlled via the inductance asymmetry $\Delta_\text{k}$ between the loop branches. The capacitances $C_\text{r}$ load the readout mode and $C_\text{c}$ couples the circuit to the readout port. The colors indicate the materials used: blue for aluminum, red for grAl and purple for aluminum covered with grAl as a result of the three-angle fabrication process. \textbf{b)} Design layout of the device. The $C_\text{c}$ pad has a skeletal shape to minimize screening currents and trapped vortices. \textbf{c)} False-coloured scanning electron micrograph of the qubit loop. By adjusting the length and width of the grAl strips the resonator frequency, coupling strength and qubit spectrum can be tuned independently. The insets show the Al/AlO$_{x}$/Al junction with an area of $A_\text{J} \approx 0.06\,\upmu\text{m}^2$ and a section of the grAl wire. The grainy texture is due to a gold film deposited for imaging. \textbf{d)} Schematics of the  microwave reflection measurement setup at 10\,mK.}

\end{figure}

To design the qubit-readout coupling we follow three design rules that will be expanded in the following paragraphs. First, we use the minimally required complexity for two electromagnetic modes, i.e. three circuit nodes.
Second, we allocate different roles to the common and differential modes to implement the resonator and qubit. The qubit mode is obtained by connecting two nodes with a JJ. The electromagnetic mode that charges these nodes out of phase inherits a large anharmonicity from the JJ, while the orthogonal in-phase mode remains harmonic.
Third, electric field coupling between the resonator and the qubit is eliminated by enforcing symmetric capacitors for the circuit nodes connecting the JJ, resulting in a permutation invariance of the capacitance matrix for these nodes.

In \figref{fig:GFQ}a we present our lumped-element circuit design which consists of three superconducting islands, i.e. circuit nodes \textbf{1}, \textbf{2} and \textbf{3}, connected via kinetic inductors made of grAl. The resulting superconducting loop, interrupted by the JJ and threaded by external flux $\Phi_{\mathrm{ext}}$, implements a generalized flux qubit (GFQ)~\cite{Yan__GFQ__2020}. 
The loop inductance $L_\text{q}$ defines the inductive energy $E_\text{L} = \Phi_0^2 / 4\pi^2\!L_\text{q}$ of the qubit, where $\Phi_0 = h/2e$ is the magnetic flux quantum.

If the circuit is symmetric with respect to the vertical symmetry line through node \textbf{3}, which means that nodes \textbf{1} and \textbf{2} have equal capacitances $C_\text{r}$ as well as equal inductances $L_\text{q}/2$, the qubit and resonator modes are electromagnetically uncoupled. The main stray capacitances of the design are discussed in \suppref{sec:circuit_model}.

The current in the readout mode splits between the qubit loop branches and in case of perfect symmetry the net shared current with the qubit mode is zero. 
To engineer qubit-readout coupling we introduce the kinetic inductive asymmetry $\Delta_\text{k} = \Delta_{\square}L_{\square}/2$ by designing different grAl inductor lengths for the two qubit inductors in the qubit loop with a total difference of $\Delta_{\square}$ squares of grAl wire with sheet inductance $L_{\square}$. In case of JJ arrays, this would translate into using different junction numbers or junction sizes for the two inductors.  
As a result, the readout mode current splits unevenly in the qubit loop and the circuit is equivalent to an inductively coupled qubit~\cite{Smith__InductiveCoupling__2016}, where $\Delta_\text{k}$ plays the role of the shared inductance. 

In contrast to capacitive~\cite{Manucharyan2009Oct, Bao__capacitively_coupled_qubit__2022} or conventional inductive~\cite{Smith__InductiveCoupling__2016, Gru__High_Impedance_Quantum_Circuits__2019} coupling of flux qubits, one circuit node is eliminated by collapsing the readout mode into the qubit loop. We would like to highlight several practical advantages of this coupling scheme. First, removing a circuit node pushes the parasitic modes to higher frequencies, improving the spectral purity of the device. Second, by coupling the resonator mode capacitively to a readout line using the capacitance $C_c$ at node \textbf{3}, the direct coupling of the qubit mode remains minimal thanks to the axial symmetry. Third, the qubit-resonator coupling can be designed to be purely inductive, loosening constraints on capacitor design, and possibly facilitating innovative flux-pumping schemes~\cite{Lu2023Sep}.

We model the~\figref{fig:GFQ}a circuit in the harmonic oscillator basis of the linearized circuit eigenmodes:

\begin{equation}
    \begin{split}
    \mathcal{H} & = \hbar \omega_\text{R} \left( \hat{a}_\text{R}^{\dagger} \hat{a}_\text{R} + \frac{1}{2} \right) + \hbar \omega_\text{Q} \left( \hat{a}_\text{Q}^{\dagger} \hat{a}_\text{Q} + \frac{1}{2} \right)\\
    & - E_\text{J} \cos \left(  
    \lambda_\text{R}  ( \hat{a}_\text{R} + \hat{a}_\text{R}^{\dagger}) 
    + \lambda_\text{Q} ( \hat{a}_\text{Q} + \hat{a}_\text{Q}^{\dagger}) 
    -  \frac{2\pi}{\Phi_0} \Phi_{\text{ext}}  
    \right) ,
    \end{split}
    \label{eq:Hamiltonian}
\end{equation} 
where $E_\text{J}$ is the Josephson energy, and $\hat{a}^{\dagger}_{\text{R},\text{Q}}$ and $\hat{a}_{\text{R},\text{Q}}$ are the bosonic creation and annihilation operators for the readout and qubit modes with eigenfrequencies $\omega_\text{R}$ and $\omega_\text{Q}$, calculated without the Josephson inductance. The qubit and readout modes are linear combinations of the common and differential modes, defined by the dimensionless coupling coefficients $\lambda_\text{R,Q}$, derived in 
\suppref{sec:circuit_model}. If we neglect the nonlinearity of the granular aluminum wire, the JJ is the sole source of nonlinearity in the system, such that the intuitive picture of a nonlinear qubit and linear readout mode is justified for $\lambda_\text{R} \ll \lambda_\text{Q}$. Note that the coupling between readout and qubit vanishes for perfect symmetry, i.e. $\lambda_\text{R} \rightarrow 0$ for $\Delta_\text{k} \rightarrow 0$.

In \figref{fig:GFQ}b we show the layout of the qubit-resonator design. A scanning electron micrograph of the qubit loop is shown in \figref{fig:GFQ}c. The qubit parameters can be tuned independently by adapting the length of the inductor $L_\text{q}$, the junction area defining $E_\text{J}$ and $C_\text{J}$, and the size of the shunt capacitor electrodes determining $C_\text{sh}$. This can be done entirely geometrically, without changing the circuit topology nor the oxidation parameters for the JJ or the grAl film.
The lateral inset in \figref{fig:GFQ}c shows a section of the grAl wire. Thanks to the relatively large kinetic inductance offered by grAl~\cite{Rotzinger__grAl__2016, Gru__Loss_Mech_grAl__2018}, it is sufficient to add a few squares of grAl film to one qubit branch to span the range from zero up to several nH of inductive asymmetry $\Delta_\text{k}$. Notably, this can be done with minimal disturbance to the geometric inductance and the capacitance matrix.

The central inset of \figref{fig:GFQ}c shows a scanning electron micrograph of the qubit Al/AlO$_{x}$/Al junction. The entire device is fabricated on a c-plane sapphire substrate in a single lithographic step using a three-angle shadow evaporation technique, similar to Ref.~\cite{Gru__High_Impedance_Quantum_Circuits__2019}. The aluminum layers (20 nm and 30 nm) are shadow evaporated to define the junction, followed by a zero-angle deposition of a 70 nm layer of grAl with resistivities between 450 $\upmu \Omega$cm and 1000 $\upmu \Omega$cm depending on the device (cf. \suppref{sec:Fabrication_method}). The sample is mounted in a modular flip-chip architecture, anchored to the base plate of a dilution-cryostat at approx. 10 mK and it is measured in reflection, as shown in \figref{fig:GFQ}d. The output signal is amplified using a dimer Josephson junction array amplifier (DJJAA)~\cite{Winkel__Amplifier__2020} operating close to the quantum noise limit.

\begin{figure*}[!htb]
    \includegraphics[width=1.0\textwidth]{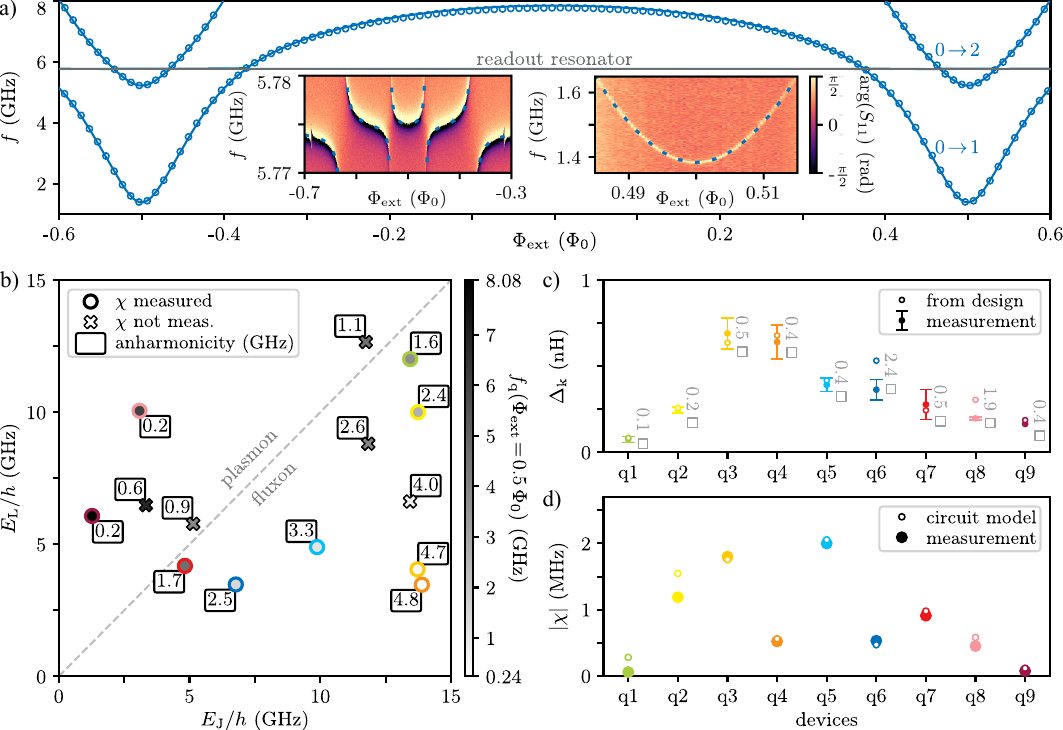}
    \caption{\label{fig:Expe_Chi_PhaseDiag} \textbf{From plasmon to fluxon: summary of measured qubit parameters.} \textbf{a)} Combined plot of typical single and two tone spectroscopy of $0\!\rightarrow\!1$ and $0\!\rightarrow\!2$ qubit transitions (blue circles) vs. flux bias $\Phi_{\text{ext}}$ of device q6 as well as the 5.77\,GHz resonance of the readout resonator (grey horizontal line). The inset on the left shows the measured phase response $\arg(S_{11})$ of the readout mode in the vicinity of the qubit-readout avoided level crossings when probing the system with a single tone. The inset on the right shows the phase response of the resonator on resonance when probing the qubit with a second tone near the qubit frequency in the vicinity of the half-flux sweet spot $\Phi_{\text{ext}} = \Phi_0/2$. The blue lines (dashed and continuous) correspond to the fitted circuit model with fit parameters $E_\text{J}$, $L_\text{q}$, $L_\text{r}$, $C_\text{J}$ and $\Delta_\text{k}$. \textbf{b)} Phase diagram $E_\text{L}$ vs. $E_\text{J}$ for the measured GFQs. The grey-scale intensity of the marker filler indicates the $0\!\rightarrow\!1$ transition frequency $f_\text{q}$ at the half-flux point, with corresponding labels indicating the anharmonicity. The diagonal grey line separates the plasmon regime on the left from the fluxon regime on the right. Devices for which the dispersive shift $\chi$ was measured (was not measured) have a circular (cross-shaped) marker. \textbf{c)} Qubit loop asymmetry $\Delta_\text{k}$ for selected devices. The filled circles indicate the values of $\Delta_\text{k}$ extracted from the joint fit of the qubit and resonator spectroscopy (cf. left inset of panel \textbf{a} and \suppref{sec:qubit_spectra}). The errorbars correspond to possible capacitive coupling arising from asymmetries $\Delta_\text{C}=\pm\! 25\,\text{aF}<\!0.01\cdot C_\text{r}$ in the capacitance matrix. The design values, shown as empty circles, are given by the product of the sheet inductance and the length difference between the qubit branches. The sheet inductance is extracted from the fitted $L_\text{q}$ and the designed number of squares in the loop. The discrepancy between the measured and design values is shown in grey labels in units of squares. The marker color assigned to each sample is consistent in all panels. \textbf{d)} Qubit state dependent dispersive shift $\chi$ at $\Phi_{\text{ext}} = \Phi_0/2$ for selected devices. Filled circles show $\chi$ values extracted from complex plane distributions of single shot measurements (cf. \suppref{sec:qubit_spectra}). Empty circles indicate the calculated $\chi$ assuming pure kinetic inductance coupling. \vspace{0.5cm} }
\end{figure*}

We measure the spectra of 14 different GFQs as a function of the external flux $\Phi_{\mathrm{ext}}$ using two-tone spectroscopy. A typical spectrum is shown in \figref{fig:Expe_Chi_PhaseDiag}a. Using the circuit Hamiltonian in \eqref{eq:Hamiltonian} we fit the qubit and resonator spectra simultaneously to obtain the circuit parameters $L_\text{r}$, $L_\text{q}$, $\Delta_\text{k}$, $C_\text{J}$ and $E_\text{J}$, while capacitors $C_\text{r}$ and $C_\text{sh}$ are inferred from finite-element simulations (see \suppref{sec:electromagnetic_simulations}). The coupling asymmetry $\Delta_\text{k}$ is determined by the width of the avoided level crossing and $L_\text{q}$, $C_\text{J}$ and $E_\text{J}$ are given by the measured qubit level structure. 

The qubit spectra can be understood in terms of universal double-well physics~\cite{Yan__GFQ__2020}, ranging from the fluxon-tunneling regime $E_\text{J} > E_\text{L}$ in which the barrier height exceeds the confining quadratic potential, to the single-well plasmon regime for $E_\text{J} < E_\text{L}$. As summarized in \figref{fig:Expe_Chi_PhaseDiag}b, the distribution of qubit frequencies and anharmonicities follow the underlying single and double-well physics: Towards the plasmon regime, the qubit frequencies increase whereas the anharmonicities decrease. Towards the fluxon regime, frequencies decrease while anharmonicities increase as expected from the exponential scaling of the qubit frequency with the barrier height~\cite{Rastelli__Phase_Slips__2013}. At half-flux bias we measure coherence times in the range of 1 $\upmu \mathrm{s}$ to 10 $\upmu \mathrm{s}$, likely limited by inductive losses in grAl as summarized in \suppref{sec:coherence_times}.

In \figref{fig:Expe_Chi_PhaseDiag}c we compare the fitted and designed coupling asymmetry $\Delta_\text{k}$. 
The qubit-readout coupling is given by the sum of designed inductive coupling via $\Delta_\text{k}$ and spurious capacitive asymmetries, which we parametrize as $\Delta_\text{C} = (C_{13} - C_{23})/2$. These asymmetries can arise from asymmetric spurious capacitances of islands \textbf{1} and \textbf{2} to ground. From finite-element simulations we estimate a maximal value of $\Delta_\text{C}=\pm 25\,\text{aF}$ due to a possible asymmetric displacement of the qubit chip with respect to ground. The corresponding uncertainties in the extraction of $\Delta_\text{k}$ are given as errorbars in \figref{fig:Expe_Chi_PhaseDiag}c.

The dispersive shift of the readout resonator is
\begin{equation*}
    \chi = \left( E_{|1,1\rangle} - E_{|0,1\rangle} \right)/h - \left(E_{|1,0\rangle} - E_{|0,0\rangle} \right)/h, 
\end{equation*} 
where $E_{|n_\text{R}, n_\text{Q}\rangle }$ is the energy level sorted by the readout ($n_\text{R}$) and qubit ($n_\text{Q}$) photon number. In \figref{fig:Expe_Chi_PhaseDiag}d we compare the measured dispersive shifts to the expected model values using the extracted $\Delta_\text{k}$ from spectroscopic measurements for nine qubits. The measured data is consistent with circuit model predictions for $\Delta_\text{C}=0$, validating the pure kinetic inductance coupling design.

\begin{figure*}[!htb]
    \includegraphics[width=1.0\textwidth]{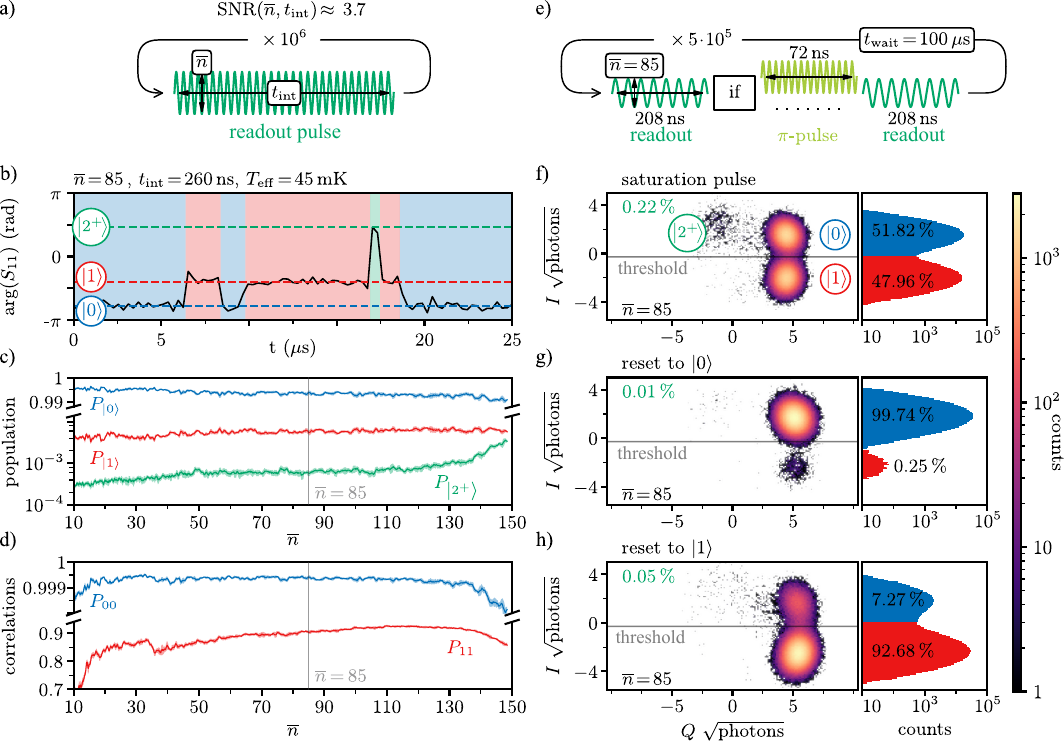}
    \caption{\label{fig:QND_Plot} \textbf{Readout fidelity and quantum state preparation.} 
    \textbf{a)} Pulse sequence for the continuous-wave measurement protocol: $10^6$ contiguous readout pulses are sent and integrated for different photon numbers $\overline{n}$. For each $\overline{n}$, the integration time $t_{\mathrm{int}}$ is adapted to keep the $\mathrm{SNR} = 3.7 \pm 0.2$. \textbf{b)} Typical quantum jump trajectory visible in the measured phase of the reflection coefficient S$_{11}$ shown in a window of $25\,\upmu \mathrm{s}$ for device q7. The qubit states are assigned using a Gaussian mixture model and indicated by the color of the background: blue ($\left|0\right\rangle$\,=\,ground), red ($\left|1\right\rangle$\,=\,excited) and green ($\left|2^{+}\!\right\rangle$\,=\,other). \textbf{c)} Measured states population vs. $\overline{n}$. Note that leakage to the $\left|2^{+}\!\right\rangle$-states accelerates for $\overline{n}\!\gtrsim\!130$. \textbf{d)} Correlation $P_{xx}$ for consecutive measurements in the ground ($x\!=\!0$) or excited ($x\!=\!1$) state vs. $\overline{n}$. The minimal integration time, 200 ns, is approximately three times larger than the resonator response time. \textbf{e)}  Pulse sequence used for active state reset. The measurement outcome of the first readout pulse is used to condition a $\pi$-pulse on the qubit. The result of the second readout is used to assess the fidelity of the reset protocol. We use $\overline{n} = 85$ and $t_{\mathrm{int}} = 208$\,ns resulting in a state separation of $\approx 6\sigma$. We repeat the sequence 5 $\times$ 10$^{5}$ times with a waiting time of $t_\text{wait}=100\,\upmu \mathrm{s}$ in between. The measured pointer state distributions for 50 \% polarization, active reset to $\ket{0}$ and $\ket{1}$ are shown in  panels \textbf{f}, \textbf{g} and \textbf{h}, respectively. The grey line is the threshold used for state assignment in the active reset protocol. The green label indicates leakage into higher states. The measurement outcomes are depicted as histograms in logarithmic scale.}

\end{figure*}   

To quantify the readout performance of our device we have performed two sets of characterization: contiguous measurement correlations and active state reset, with pulse sequences detailed in \figref{fig:QND_Plot}a \& e, respectively. We define the signal to noise ratio (SNR) in the $I$-$Q$ plane of the readout mode as the distance between pointer states corresponding to qubit in $|0\rangle$ and $|1\rangle$, divided by the sum of their standard deviations. In all experiments we fix SNR $\approx 3.7$, which is obtained by adjusting the integration times $t_{\mathrm{int}} \in (1600$, $208$)~\,ns depending on the different average photon numbers $\overline{n} \in (10, 150)$ in the resonator.

In \figref{fig:QND_Plot}b we show an example section of a contiguously measured quantum jump trace for GFQ device q7. By applying a Gaussian mixture model to quantum jump traces with $10^6$ points for a given $\overline{n}$, we extract qubit populations in $|0\rangle$, $|1\rangle$ and $|2^{+}\rangle$ (see \figref{fig:QND_Plot}c) and correlations $P_{\mathrm{00}}$ and $P_{\mathrm{11}}$ for two successive measurements in the ground and excited state (see \figref{fig:QND_Plot}d), respectively. The correlations $P_{00}$ and $P_{11}$ serve as a measure of qubit readout fidelity, particularly useful to assess quantum demolition effects introduced by the readout drive. Similarly to Ref.~\cite{Gusenkova__High_photon_number__2021}, the resilience of the grAl GFQ to readout-induced leakage ~\cite{Dumas__unified_ionization__2024, Shillito__Dynamics_Ionization__2022, Cohen__Reminiscence_of_Classical_Chaos_in_Driven_Transmons__2023} is illustrated by the fact that up to $\overline{n} \approx 100$, the qubit populations remain approximately constant, corresponding to an effective temperature of about 40 mK to 45 mK and residual excitations outside of the computational subspace remain below 0.1\,\%. 

Within the qubit subspace, we observe a significant difference in the correlation of successive readout outcomes when the qubit is in the ground or excited state. Qubit measurements in the ground state are highly correlated, with $P_{\mathrm{00}} > 99.9\,\%$ for a broad range of readout powers. In contrast, we find that $P_{\mathrm{11}}$ depends on the readout strength, with $P_{\mathrm{11}} \geq 90\,\%$ for $\overline{n} \in (75,140)$. The difference between the measured $P_{11}$ and perfect correlation can be accounted for by summing three contributions: Energy decay during the measurement reduces $P_{11}$ by $1-\exp(-t_{\mathrm{int}}/T_1)$, which for $t_{\mathrm{int}} = 352 \, \mathrm{ns}$ can be as high as 6 \% given the measured $T_1  = 8.0 \pm 2.4 \, \upmu \mathrm{s}$ for device q7. Second, the qubit spectral shift and broadening induced by the readout tone will change the dissipative environment of the qubit~\cite{Thorbeck__Zeno_IBM__2024} and might accelerate the relaxation from the excited to the ground state. The third contribution comes from demolishing effects activated when increasing $\overline{n}$~\cite{Walter__QD_transmon__2017}, such as leakage outside of the qubit subspace~\cite{Dumas__unified_ionization__2024}. The second and third contributions, which sum up to give at least $ 4 \%$ of the $P_{11}$ infidelity, provide a measure for the performance of the qubit-readout coupling scheme and motivate future research efforts.

We implement active state preparation starting from the thermal state of the qubit by playing a conditional $\pi$-pulse. The threshold to discriminate states $|0\rangle$ and $|1\rangle$ is determined by measuring the $I$-$Q$ plane distributions after a saturation pulse, as shown in \figref{fig:QND_Plot}f. Using  $\overline{n} = 85$, the fidelities to reset the qubit to its ground and excited state read $P^{\mathrm{active}}_{\mathrm{0}} = 99.7 \%$ and $P^{\mathrm{active}}_{\mathrm{1}} = 92.7 \%$, respectively (cf. \figref{fig:QND_Plot}g \& h).
In the error budget for quantum state preparation, the fidelity of the $\pi$-pulse of $>$99 \% (cf. \suppref{sec:pulse}) is a negligible contribution compared to the decay during readout and quantum demolition effects. The measured performance for our GFQ devices are similar to results reported for fluxoniums and transmons~\cite{Gusenkova__High_photon_number__2021, Tholen__Active_Reset__2022, Riste__Active_Reset__2012} but below state-of-the-art fidelity reaching $99\,\%$~\cite{Sunada__Active_Reset__2022}. Currently, the main limitation for the readout performance is the energy relaxation time of the qubit, which can be significantly improved via material and design optimization~\cite{Siddiqi_High_Coherence_Materials__2021}. 

We have demonstrated dispersive coupling between a harmonic mode and a generalized flux qubit consisting of a single junction shunted by a granular aluminum inductor. By embedding the harmonic readout mode into the high kinetic inductance loop of the flux qubit we implement a mechanism conceptually equivalent to inductive coupling, where the loop asymmetry is equivalent to the shared inductance. 
We validate the kinetic inductance coupling concept by comparing the spectra of 14 devices obtained via two-tone spectroscopy to a model including parasitic capacitances.
We assess the suitability of the coupling mechanism for dispersive readout by performing quantum non-demolition readout with $> 90\, \%$ active state preparation fidelity and less than 0.1\,\% leakage outside the qubit computational space. Thanks to its ability to provide a local qubit-resonator interaction unaffected by on-chip capacitors, we believe that the minimalist qubit-resonator design presented here will provide an advantageous avenue for up-scaling superconducting quantum devices.

\section*{Acknowledgements}
 We are grateful to L. Radtke and S. Diewald for technical assistance. We acknowledge funding from the European Commission (FET-Open AVaQus GA 899561). Facilities use was supported by the KIT Nanostructure Service Laboratory. We acknowledge the measurement software framework qKit. The authors acknowledge support by the state of Baden-Württemberg through bwHPC. M.S., P.P, N.G. and T.R. acknowledge support from the German Ministry of Education and Research (BMBF) within the project GEQCOS (FKZ: 13N15683). D.B. and M.F. acknowledge funding from the German Federal Ministry of Education and Research (BMBF) within the project QSolid (FKZ: 13N16151). N.Z. acknowledges funding from the Deutsche Forschungsgemeinschaft (DFG – German Research Foundation) under project number 450396347 (GeHoldeQED). S.Gü., D.R. and W.W. acknowledge support from the Leibniz award WE 4458-5.

\bibliography{v2_aipsamp}

\newpage
\onecolumngrid

\setcounter{equation}{0}
\setcounter{figure}{0}
\setcounter{table}{0}
\makeatletter
\renewcommand{\theequation}{S\arabic{equation}}
\renewcommand{\thefigure}{S\arabic{figure}}
\renewcommand{\thetable}{S\arabic{table}}

\newpage
\section*{Appendix}

\subsection{\label{sec:circuit_model} Circuit Models}

\begin{figure*}[!htb]
    \includegraphics[width=1.0\textwidth]{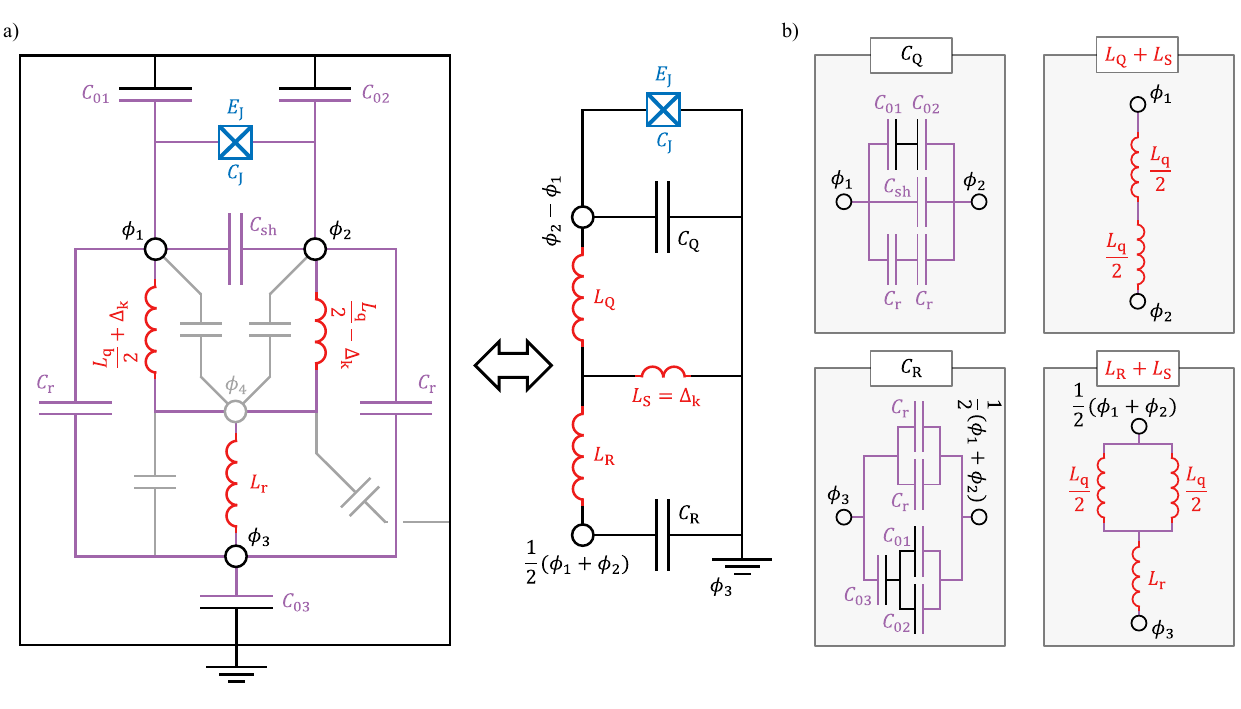}
    \caption{\label{fig:full_circuit} \textbf{Extended and idealized circuit model.} \textbf{a)} In addition to islands \textbf{1} to \textbf{3} discussed in the main text, the extended circuit model describes all four islands (black empty circles) including island \textbf{4}, which is marked in grey. Blue, purple and red colors indicate the materials used, consistent with \figref{fig:GFQ} in the main text. Neglecting the influence of island \textbf{4} and assuming symmetric capacitances, the extended circuit is equivalent to an idealized inductively coupled circuit via the shared inductance $L_\text{S} = \Delta_\text{k}$, in which $L_{\text{Q},\text{R}}$ and $C_{\text{Q},\text{R}}$ are the equivalent inductance and capacitance of the qubit and resonator, respectively.
    \textbf{b)} The four panels depict the transformations required to obtain the equivalent circuit branch components $C_\text{Q}$, $C_\text{R}$, $L_\text{Q}$ and $L_\text{R}$.
    } 
\end{figure*}  

The circuit is modeled in terms of the flux node variables $\Vec{\phi}^{\intercal} = (\phi_1, \phi_2, \phi_3, \phi_4)$ and illustrated in \figref{fig:full_circuit}. Following the canonical circuit quantization procedure, the Lagrangian reads

\begin{equation}
    \begin{split}
    \mathcal{L} & = \frac{1}{2} \dot{\vec{\phi}}^{\intercal} \mathbf{C} \dot{\vec{\phi}} \, + \, \frac{1}{2} \vec{\phi}^{\intercal} \mathbf{L}^{-1} \vec{\phi} \, + \, E_\text{J} \cos \left( \frac{2\pi}{\Phi_0} (\phi_2 \, - \, \phi_1 \, - \, \Phi_{\mathrm{ext}} ) \right) \\
    & = \mathcal{L}_{\mathrm{lin}} \, + \, E_\text{J} \cos \left( \frac{2\pi}{\Phi_0} (\phi_2 \, - \, \phi_1 \, - \, \Phi_{\mathrm{ext}} ) \right)
    .
    \end{split}
    \label{eq:Lagrangian}
\end{equation} where $\mathcal{L}_{\mathrm{lin}}$ is the linearized circuit Lagrangian without the junction cosine potential, $\Phi_{\mathrm{ext}}$ is the external flux through the qubit loop, $\mathbf{C}$ is the capacitance matrix and $\mathbf{L}^{-1}$ the inverse inductance matrix, given by

\begin{equation*}
    \mathbf{C} =
    \begin{pmatrix}
        C_{11}+C_{\text{J}} & -C_{12} - C_\text{J} & -C_{13} & -C_{14}\\
        
        -C_{12} - C_\text{J} & C_{22} + C_{\text{J}} & -C_{23} & -C_{24}\\
        
        -C_{13} & -C_{23} & C_{33} & -C_{34}\\
        
        -C_{14} & -C_{24} & -C_{34} & C_{44}\\   
    \end{pmatrix}
    \label{C_matrix}
\end{equation*}

and
\begin{equation*}
    \mathbf{L}^{-1} =
    \begin{pmatrix}
        \frac{1}{L_\text{q}/2+\Delta} & 0 & 0 & \frac{-1}{L_\text{q}/2+\Delta} \\
        0 & \frac{1}{L_\text{q}/2-\Delta} & 0 & \frac{-1}{L_\text{q}/2-\Delta} \\
        0 & 0 & \frac{1}{L_\text{r}} & \frac{-1}{L_\text{r}}\\
        \frac{-1}{L_\text{q}/2+\Delta} & \frac{-1}{L_\text{q}/2-\Delta} & \frac{-1}{L_\text{r}} & \frac{1}{L_\text{q}/2+\Delta}+\frac{1}{L_\text{q}/2-\Delta}+\frac{1}{L_\text{r}}  \\
    \end{pmatrix}.
    \label{L_matrix_m1}
\end{equation*}
The capacitances $C_{ij}$ are obtained by finite-element simulations and include the capacitances between the islands on the chip and the sample holder (ground) $C_{i0}$. The junction parallel plate capacitance $C_\text{J}$ is excluded from the simulations and enters as an additional capacitance between nodes \textbf{1} and \textbf{2}. A possible misalignment of the floating chip with respect to the symmetric sample holder environment gives the capacitive asymmetry 
\begin{equation*}
    \Delta_\text{C} = \frac{C_{13} - C_{23}}{2}.
\end{equation*}

\subsubsection*{Extended circuit model}

We start with a coordinate transformation from $\Vec{\phi}$ to $\vec{x} = \mathbf{C}^{1/2} \vec{\phi}$, in which the linearized Lagrangian reads
\begin{equation*}
\mathcal{L}_{\mathrm{lin}} = \frac{1}{2} \dot{\vec{x}}^{\intercal}\dot{\vec{x}}
\, - \frac{1}{2} \vec{x}^{\intercal} \mathbf{C}^{-\frac{1}{2}} \mathbf{L}^{-1}\mathbf{C}^{-\frac{1}{2}} \vec{x}.
\end{equation*}
The bare normal modes $\vec{\eta}_j$ are then found by solving the eigenvalue problem
\begin{equation*}
  \mathbf{C}^{-\frac{1}{2}} \mathbf{L}^{-1}\mathbf{C}^{-\frac{1}{2}} \vec{\eta}_j = \omega^2_j \vec{\eta}_j
\end{equation*}
via the transformation matrix $\mathbf{S} = (\vec{\eta}_1, \vec{\eta}_2, \vec{\eta}_3, \vec{\eta}_4 )$, such that the eigenbasis is written as $\vec{\eta} = (\eta_1, \eta_2, \eta_3, \eta_4)^{\intercal} = \mathbf{S}^{\intercal} \vec{x}$. By introducing the conjugate momenta $p_i = \frac{\partial \mathcal{L}_{\mathrm{lin}}}{\partial \dot{\eta}_i} $ we obtain the diagonal linearized Hamiltonian
\begin{equation*}
\mathcal{H}_{\mathrm{lin}} = \sum_i p_i \dot{\eta_i} -\mathcal{L}_{\mathrm{lin}} = \frac{1}{2}\sum_i \left( p_i^2 + 
\omega_i^2 \eta_i^2 \right).
\end{equation*}

The eigenmodes are sorted with respect to their even and odd amplitudes across the junction, meaning that the qubit mode is defined as $\eta_\text{Q} = \eta_1$ with $|\mathbf{S_{21}} - \mathbf{S_{11}}| > |\mathbf{S_{2j}} - \mathbf{S_{1j}}| \, \forall j \neq 1$ and the readout mode as $\eta_\text{R} = \eta_2$ with $|\mathbf{S_{22}} + \mathbf{S_{12}}| > |\mathbf{S_{2j}} + \mathbf{S_{1j}}| \, \forall j \neq 2$. The back transformation from the harmonic oscillator (H.O) basis to the original flux node variables then reads

\begin{equation*}
    \vec{\phi} =  \underbrace{\mathbf{C}^{-\frac{1}{2}} \mathbf{S}}_{\mathbf{S}'}  \Vec{\eta}.
\end{equation*}

The common and differential modes can then be expressed in terms of qubit and readout modes as

\begin{equation}
    \begin{split}
    \phi_2 \, - \, \phi_1 & \approx  \left( \mathbf{S}'_{2\text{Q}} -\mathbf{S}'_{1\text{Q}} \right)   \eta_\text{Q}
    \, + \,
    \left(  \mathbf{S}'_{2\text{R}} -\mathbf{S}'_{1\text{R}} \right) \eta_\text{R}  \\
    \phi_2 \, + \, \phi_1 & \approx  \left( \mathbf{S}'_{2\text{Q}} +\mathbf{S}'_{1\text{Q}} \right)   \eta_\text{Q}
    \, + \,
    \left(  \mathbf{S}'_{2\text{R}} +\mathbf{S}'_{1\text{R}} \right) \eta_\text{R} 
    \end{split}
    \label{eq:phi1_phi2}
\end{equation}
where we neglect the presence of the  zero mode and the higher frequency mode associated with island \textbf{4}. This simplification~\cite{Rymarz__consistent_quantization__2023} is justified in our case because the mode associated with island $\textbf{4}$, at $\approx 30 \, \mathrm{GHz}$, is sufficiently detuned from the spectrum of the device to give negligible corrections.

In the H.O. basis, coupling is solely mediated by the junction cosine term driven by the differential mode. To quantify it, we express the cosine argument in Eq. \ref{eq:Lagrangian}
via the above transformation and quantize the qubit-readout system and apply the canonical quantization procedure to the qubit and readout variables
\begin{equation*}
    \begin{split}
    \eta_{\text{R},\text{Q}} & = \sqrt{\frac{\hbar}{2\omega_{\text{R},\text{Q}}}} \left( \hat{a}_{\text{R},\text{Q}} + \hat{a}_{\text{R},\text{Q}}^{\dagger} \right) \\
    p_{\text{R},\text{Q}} & = i\sqrt{\frac{\hbar\omega_{\text{R},\text{Q}}}{2}} \left( \hat{a}_{\text{R},\text{Q}}^{\dagger} - \hat{a}_{\text{R},\text{Q}} \right).\
    \label{eq:quantization}
    \end{split}
\end{equation*}

Using this procedure we obtain the main text Hamiltonian

\begin{equation*}
    \begin{split}
    \mathcal{H} &=  \hbar \omega_\text{R} \left( \hat{a}_\text{R}^{\dagger} \hat{a}_\text{R} + \frac{1}{2} \right) + \hbar \omega_\text{Q} \left( \hat{a}_\text{Q}^{\dagger} \hat{a}_\text{Q} + \frac{1}{2} \right)\\
    & - E_\text{J} \cos \left(  \lambda_\text{R} ( \hat{a}_\text{R} + \hat{a}_\text{R}^{\dagger}) + \lambda_\text{Q} ( \hat{a}_\text{Q} + \hat{a}_\text{Q}^{\dagger}) 
    - \frac{2\pi}{\Phi_0} \Phi_{\text{ext}}  \right)
    \end{split}
\end{equation*}
with 
\begin{align*}
\lambda_\text{R}  &= \frac{2\pi}{\Phi_0}\sqrt{\frac{\hbar}{2\omega_\text{R}}}\left( \mathbf{S}'_{2\text{R}} -\mathbf{S}'_{1\text{R}} \right) \qquad \qquad 
\lambda_\text{Q} = \frac{2\pi}{\Phi_0}\sqrt{\frac{\hbar}{2\omega_\text{Q}}}\left( \mathbf{S}'_{2\text{Q}} -\mathbf{S}'_{1\text{Q}} \right). 
\end{align*}
To obtain the qubit-resonator spectrum, we numerically diagonalize the Hamiltonian in the photon number basis using 15 and 30 basis states for the resonator and qubit, respectively.

\subsubsection*{Idealized circuit model}
While the procedure discussed in the previous section is sufficient to calculate the qubit-resonator spectrum, in order to gain intuition on the current and future designs, we find it instructive to also discuss the idealized model for a symmetric capacitive environment with $\Delta_\text{C} = 0$. Below, we simplify the linearized extended circuit model in terms of the flux nodes $\Vec{\phi}^{\intercal} = (\phi_1, \phi_2)$, c.f. \eqref{eq:phi1_phi2}. Neglecting the capacitances to the center node \textbf{4}, we can eliminate this inactive node via Kirchhoff's rule of conserved currents 
\begin{equation*}
    \frac{\phi_4 - \phi_3}{L_\text{r}} = \frac{\phi_2 - \phi_4}{\frac{L_\text{q}}{2} - \Delta_\text{k}} + \frac{\phi_1 - \phi_4}{\frac{L_\text{q}}{2} + \Delta_\text{k}}.
\end{equation*}
Further removing the zero mode with respect to the sample holder (ground), we can gauge $\phi_3 = 0$ as the new reference potential and write
\begin{equation}
    \phi_4 = \Sigma_\text{L}^{-1} \left(  L_\text{r} \left(\frac{L_\text{q}}{2} + \Delta_\text{k} \right) \phi_2 + L_\text{r} \left(\frac{L_\text{q}}{2} - \Delta_\text{k} \right) \phi_1\right)
    \label{eq:Kirchhoff}
\end{equation}
where we define
\begin{equation*}
    \Sigma_\text{L} = L_\text{r} L_\text{q} + \frac{L_\text{q}^2}{4} - \Delta_\text{k}^2. 
\end{equation*}

The inductive contribution to the Lagrangian is
\begin{equation*}
    U = \frac{1}{2L_\text{r}} \phi_4^2 + \frac{1}{2\left( \frac{L_\text{q}}{2} + \Delta_\text{k}\right) } (\phi_1 - \phi_4)^2 + \frac{1}{2\left( \frac{L_\text{q}}{2} - \Delta_\text{k}\right) } (\phi_2 - \phi_4)^2.
\end{equation*}
Using Eq. \ref{eq:Kirchhoff},
\begin{equation*}
    U = \frac{1}{2\Sigma_\text{L}} \left(
    \left( L_\text{r} +\left( \frac{L_\text{q}}{2} - \Delta_\text{k}\right) \right)\phi_1^2 +
    \left( L_\text{r} +\left( \frac{L_\text{q}}{2} + \Delta_\text{k}\right) \right)\phi_2^2
    - 2L_\text{r}\phi_1 \phi_2
    \right),
\end{equation*}
which can be rewritten in matrix form as $U = \frac{1}{2} \Vec{\phi}^{\intercal} \mathbf{L}^{-1} \Vec{\phi}$, with the inverse inductance matrix
\begin{equation*}
    \mathbf{L}^{-1} = \frac{1}{\Sigma_\text{L}} 
      \begin{pmatrix}
        L_\text{r} + \frac{L_\text{q}}{2} - \Delta_\text{k}  & -L_\text{r} \\
        -L_\text{r} & L_\text{r} + \frac{L_\text{q}}{2} + \Delta_\text{k} \\
    \end{pmatrix}.
\end{equation*}

\begin{figure*}[!b]
    \includegraphics[width=0.714\textwidth]{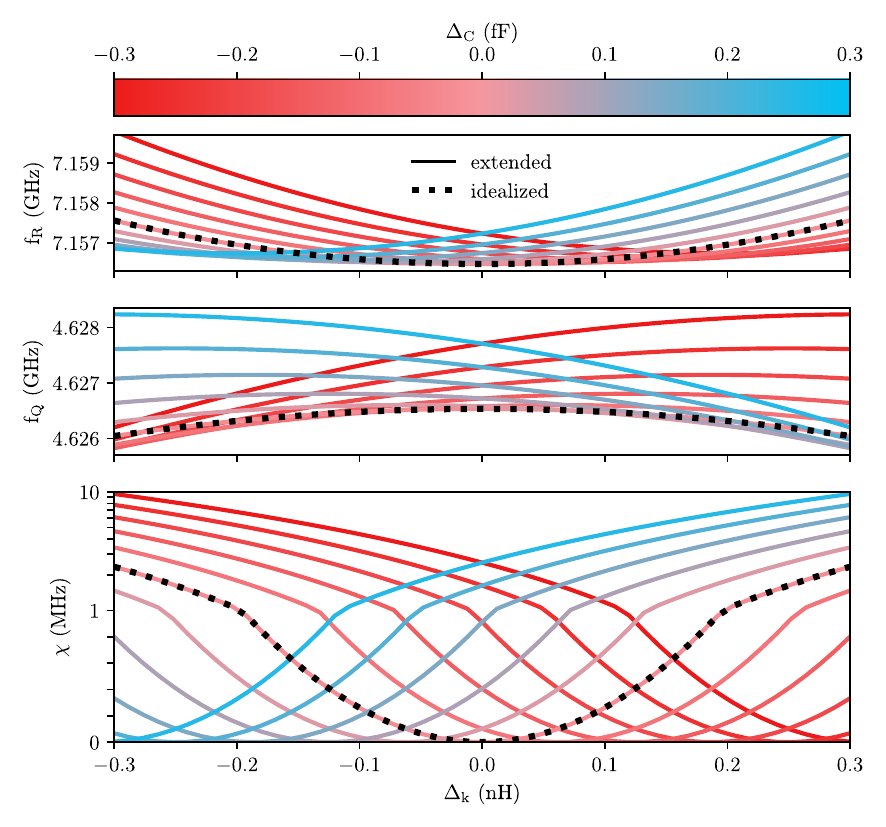}
    \caption{\label{fig:chi_fr_fq} \textbf{Numerical comparison between the extended and idealized circuit model and the effect of capacitive and inductive asymmetries.} The three plots calculated at $\phi_{\mathrm{ext}} = \frac{\Phi_0}{2}$ as a function of inductive asymmetry $\Delta_\text{k}$ show, from top to bottom, the readout and qubit frequencies $f_\text{R}$ and $f_\text{Q}$, respectively, and the dispersive shift. For $\Delta_\text{C} = 0$, the idealized model (dotted line) gives the same results as the extended model. For finite $\Delta_\text{C}$, the capacitive coupling adds to the inductive coupling, introducing a shift in the curves. Notice that a fF of capacitive asymmetry roughly equals a nH of inductive asymmetry in terms of the resulting dispersive shift. Technologically relevant dispersive shifts on the order of MHz can be implemented with 100's of aF or 100's of pH asymmetry.
    }
\end{figure*}  

To account for the contribution of the stray capacitances to ground, we treat the ground as a free node with conserved charge with respect to the floating islands and eliminate it from the Lagrangian by writing 
\begin{equation*}
    \dot{\phi}_0 = \frac{C_{\text{J}0}}{C_{30} + 2C_{\text{J}0}}(\dot{\phi}_1 + \dot{\phi}_2),
\end{equation*}
where $C_{\text{J}0} = C_{10} = C_{20}$ is the junction electrode capacitance to ground. The capacitive contribution to the Lagrangian then is

\begin{equation*}
    T = \frac{1}{2} \left( C_\text{r} + C_{\mathrm{sh}} + C_{\text{J}0} - \frac{C_{\text{J}0}^2}{C_{30} + 2C_{\text{J}0}} \right) \left( \dot{\phi}_1^2 + \dot{\phi}_2^2 \right) 
    - \left( C_{\mathrm{sh}} + \frac{C_{\text{J}0}^2}{C_{30} + 2C_{\text{J}0}} \right)  \dot{\phi}_1 \dot{\phi}_2
\end{equation*}
which can be rewritten in matrix form as $T = \frac{1}{2} \dot{\vec{\phi}}^{\intercal} \mathbf{C} \dot{\vec{\phi}}$, with the capacitance matrix
\begin{equation*}
    \mathbf{C} =  
      \begin{pmatrix}
        C_\text{r} + C_{\mathrm{sh}} + C_{\text{J}0} - \frac{C_{\text{J}0}^2}{C_{30} + 2C_{\text{J}0}}   & -C_{\mathrm{sh}} - \frac{C_{\text{J}0}^2}{C_{30} + 2C_{\text{J}0}} \\
        -C_{\mathrm{sh}} - \frac{C_{\text{J}0}^2}{C_{30} + 2C_{\text{J}0}} &  C_\text{r} + C_{\mathrm{sh}} + C_{\text{J}0} - \frac{C_{\text{J}0}^2}{C_{30} + 2C_{\text{J}0}} 
    \end{pmatrix}.
\end{equation*}

For $\Delta_\text{k} = 0$, a transformation that diagonalizes both $\mathbf{C}$ and  $\mathbf{L}^{-1}$ simultaneously is 
\begin{align*}
    \phi_\text{Q} &= \phi_2 - \phi_1 \\
    \phi_\text{R} &= \frac{1}{2} (\phi_1 + \phi_2),
\end{align*}
with the corresponding basis
\begin{equation*}
    \vec{\phi}^* =  
    \begin{pmatrix}
        \phi_\text{R}  \\
        \phi_\text{Q} \\
    \end{pmatrix}.
\end{equation*}

Applying the above transformation for finite $\Delta_\text{k}$ yields the matrices in the new basis
\begin{equation*}
    \mathbf{{L}^{*}}^{-1} = \frac{1}{\Sigma_\text{L}} 
    \begin{pmatrix}
        L_\text{q}   & -\Delta_\text{k} \\
        - \Delta_\text{k} & L_\text{r} + \frac{L_\text{q}}{4}\\
    \end{pmatrix}
\end{equation*}
and
\begin{equation*}
    \mathbf{C}^{*} =  
    \begin{pmatrix}
        2C_\text{r} + \left(\frac{1}{2C_{\text{J}0}} + \frac{1}{C_{30}} \right)^{-1}  & 0 \\
        0 & C_\text{r}/2 + C_{\mathrm{sh}} + C_{\text{J}0}/2 \\
    \end{pmatrix}.
\end{equation*}

Finally, we define the qubit (Q) and readout (R) effective parameters, c.f \figref{fig:full_circuit}b, as
\begin{equation*}
    \begin{split}
    L_\text{Q} & = L_\text{q} - \Delta_\text{k}\\
    L_\text{R} & = L_\text{r} + \frac{L_\text{q}}{4} - \Delta_\text{k} \\
    L_\text{S} & = \Delta_\text{k} \\
    C_\text{R} & = 2C_\text{r} + \left(\frac{1}{2C_{\text{J}0}} + \frac{1}{C_{30}} \right)^{-1} \\
    C_\text{Q} & = C_\text{r}/2 + C_{\mathrm{sh}} + C_{\text{J}0}/2,\\
    \end{split}
\end{equation*}
such that the matrices can be rewritten as
\begin{equation*}
    \mathbf{{L}^{*}}^{-1} = \frac{1}{L_\text{R} L_\text{Q} + L_\text{R} L_\text{S} + L_\text{Q} L_\text{S}} 
    \begin{pmatrix}
        L_\text{Q} + L_\text{S}  & -L_\text{S} \\
        -L_\text{S} & L_\text{R} + L_\text{S}\\
    \end{pmatrix}
\end{equation*}
and
\begin{equation*}
    \mathbf{C}^{*} =  
    \begin{pmatrix}
        C_\text{R}  & 0 \\
        0 & C_\text{Q}\\
    \end{pmatrix},
\end{equation*}
as expected from the idealized inductively coupled circuit \figref{fig:full_circuit}b which is similar to Ref.~\cite{Gusenkova__High_photon_number__2021}. A numerical comparison between the idealized circuit and the extended circuit is shown in \figref{fig:chi_fr_fq}.

\newpage

\subsection{\label{sec:Fabrication_method} Fabrication}

\begin{figure*}[!htb]
    \includegraphics[width=1.0\textwidth]{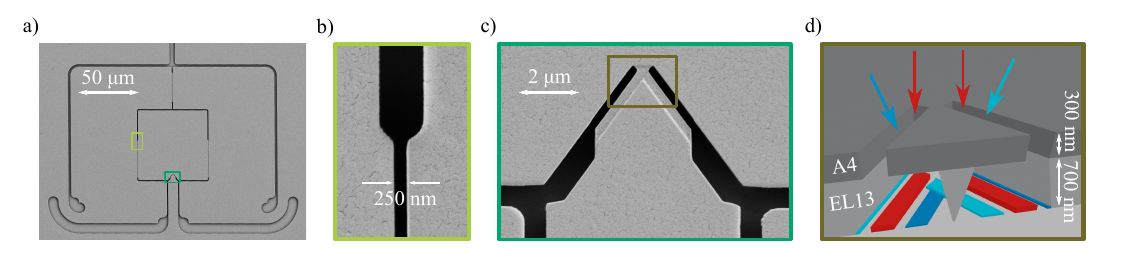}   
    \caption{\label{fig:Chip_fabrication} \textbf{Fabrication using a three-angle evaporation process.  a)} Scanning electron microscope image of the device during fabrication after e-beam patterning and development of the two-layer resist stack. The surface is covered by a gold film for imaging. \textbf{b)} Zoom-in on the pattern defining the connection of the thin grAl inductor to the Al-shunted islands. \textbf{c)} Zoom-in on the cross-junction pattern used to fabricate the JJ. In the brown-highlighted area, an undercut is used to separate the Al depositions from the zero-angle grAl deposition. \textbf{d)} Sketch of the Dolan-bridge and the depositions of the three different layers of Al (blue) and grAl (red). The undercut in the EL-13 (700\,nm) + PMMA-A4 (300\,nm) resist stack allows to create the entire device in a single three-angle evaporation step with subsequent evaporation angles $-\alpha, 0^{\circ}, \alpha$, where $\alpha \in [26^{\circ}, 30^{\circ}]$, depending on the sample.}  
\end{figure*}  

The devices are fabricated on a double-side polished c-plane 2\,inch sapphire wafer with a thickness of 330\,$\upmu$m. A two level resist stack MMA(8.5)MAA EL 13 (700\,nm thick layer) / 950 PMMA A4 (300\,nm thick layer) is spin-coated onto the wafer. An approx. $5 ~\mathrm{nm}$ thick gold layer is sputtered onto the wafer before  e-beam writing. A 50\,keV e-beam writer is used to pattern the mask. After e-beam exposure a 15\,\% Lugol solution is used to remove the gold. The resist stack is developed in an IPA/H$_2$0 3:1 solution at 6$^{\circ}$C for 90\,s. Images of the resulting mask are shown in \figref{fig:Chip_fabrication}a to \figref{fig:Chip_fabrication}c. A Niemeyer-Dolan bridge shown in \figref{fig:Chip_fabrication}c \& d is used to fabricate the Josephson junction. A commercial, controlled-angle e-beam evaporation machine (Plassys MEB 550S${}^\text{TM}$) is used to implement the three-angle evaporation process as shown in \figref{fig:Chip_fabrication}d. The evaporation steps are detailed in the paragraph below. For the mask liftoff, the wafer is immersed for two hours in a 60$^{\circ}$C Acetone bath, which is stirred for 2 minutes every 20 minutes.\\

\subsubsection*{Evaporation procedure}

The evaporation machine has two chambers. The first chamber is used as loadlock and for oxidation, while the second UHV chamber is used for evaporation. In detail the following steps are used during evaporation:
\begin{itemize}
  \item Pump the loadlock for a minimum of 2 hours until the pressure is smaller than $5\cdot10^{-7}$ mbar.
  \item Plasma cleaning process at 0$^{\circ}$ angle (Kaufman source parameters: 200 V beam voltage, 10 mA, 10 sccm O${}_2$, 5 sccm Ar)
  \item Titanium evaporation with closed shutter (10\,s with 0.2\,nm/s)
  \item Aluminium evaporation at -$\alpha$ with open shutter (Al crucible 1,  20\,nm with 1\,nm/s)
  \item Static oxidation (pure O${}_2$) of Josephson junction at 50\,mbar for 4\,min (+20\,s to linearly increase the pressure to 50\,mbar)
  \item Aluminium evaporation at $\alpha$ with open shutter (Al crucible 1, 30\,nm with 1\,nm/s)
  \item Argon milling process at 0$^{\circ}$ angle (Kaufman source parameters: 400 V beam voltage, 15 mA, 0 sccm O${}_2$, 4 sccm Ar)
  \item Regulate aluminium evaporation at 0$^{\circ}$ angle to 2\,nm/s (Al crucible 2)
  \item Regulate oxygen flow to 9.4 sccm and start planetary rotation with 5 rpm 
  \item Open shutter and evaporate for $\sim$ 35\,s (corresponds to 70\,nm of grAl)
  \item Close shutter, terminate oxygen flow, stop planetary rotation, ramp down aluminium evaporation rate
\end{itemize}

Using the process described above grAl film resistivities at room temperature vary between 450\,$\upmu\Omega$cm and 1000\,$\upmu\Omega$cm as can be seen in Table~\ref{tab:Sheet_Resistance_grAl}. 

\begin{table*}[!htb]\centering
    \begin{tabular}{|c||c|c|c|}
        \hline
        Date & $\rho$ [$\upmu\Omega$cm] \\
        \hline
        \hline
         Sep 2022 & 620 \\
        \hline
         Sep 2022 & 660 \\
        \hline
         Nov 2022 & 840 \\
        \hline
         Nov 2022 & 940 \\
        \hline
         Nov 2022 & 640 \\
        \hline
         Dec 2022 & 880 \\
        \hline
         Dec 2022 & 1000 \\
        \hline
         Feb 2023 & 720  \\
        \hline
         Mar 2023 & 530 \\
        \hline
         Sep 2023 & 520 \\
        \hline
         Dec 2023 & 470 \\
        \hline
  \end{tabular}
  \caption{\textbf{Resistivity of different grAl films at room temperature as a function of deposition date.} We use the same evaporation procedure for all grAl depositions and the thickness of the grAl film is 70\,nm.  The variability of grAl resistivity over timescales of months could be due to changes in the level of aluminum in the crucible, humidity or other cleanroom conditions.
  \label{tab:Sheet_Resistance_grAl}}
\end{table*}

\subsection{\label{sec:qubit_spectra} Measured and fitted spectra}

The circuit model described in \suppref{sec:circuit_model} is fitted to the qubit spectra using a minimization method based on the python module scipy.minimize. In \figref{supp:fig__all_qubit_spectra} the extracted data points are shown for each qubit. The resulting fit parameters $L_\text{q}$, $L_\text{r}$, $\Delta_\text{k}$, $E_\text{J}$ and $C_\text{J}$ are listed in Table \ref{supp: tab__fit_parameter__chi} for each qubit. The measured dispersive shift $\chi$ is compared to its calculated value in the columns 7 and 8 of Table \ref{supp: tab__fit_parameter__chi}.

 \begin{table*}[h]\centering   
    \begin{tabular}{|c||c|c|c|c|c||c|c|}
        \hline
        device & $L_\text{r}$\,[nH] & $L_\text{q}$\,[nH] & $\Delta_\text{k}$\,[nH] & $E_\text{J}$\,[GHz] & $C_\text{J}$\,[fF] & $\chi_\text{fit}$\,[MHz] & $\chi_\text{meas}$\,[MHz] \\
        \hline
        \hline
        q1 & $15.03$ & $13.61$ & $-0.07$ & $13.44$  & $4.82$ & $0.28$ & $0.06$ \\
        \hline
        q2 & $6.53$ & $16.34$ & $-0.24$ & $13.74$  & $6.01$ & $1.55$ & $1.19$ \\
        \hline
        q3 & $29.36$ & $40.35$ & $0.69$ & $13.72$  & $3.78$ & $-1.75$ & $-1.80$ \\
        \hline
        q4 & $30.14$ & $47.07$ & $0.64$ & $13.89$  & $5.01$ & $-0.56$ & $-0.52$ \\
        \hline
        q5 & $11.90$ & $33.38$ & $0.39$ & $9.88$  & $4.02$ & $2.05$ & $2.00$ \\
        \hline
        q6 & $7.40$ & $46.96$ & $0.36$ & $6.77$  & $3.34$ & $0.47$ & $0.53$ \\
        \hline
        q7 & $11.73$ & $39.06$ & $0.28$ & $4.83$  & $1.85$ & $0.98$ & $0.91$ \\
        \hline
        q8 & $0.55$ & $16.25$ & $0.20$ & $3.09$  & $2.58$ & $0.58$ & $0.45$ \\
        \hline
        q9 & $18.89$ & $26.91$ & $0.17$ & $1.28$  & $1.86$ & $0.11$ & $0.07$ \\
        \hline
        \hline
        q10 & $11.10$ & $18.55$ & $0.22$ & $11.84$  & $2.13$ & - & - \\
        \hline
        q11 & $9.82$ & $12.91$ & $-0.31$ & $11.75$  & $4.97$ & - & - \\
        \hline
        q12 & $6.54$ & $24.68$ & $-0.30$ & $13.44$  & $4.97$ & - & - \\
        \hline
        q13  & $7.60$ & $28.28$ & $0.86$ & $5.14$  & $4.37$ & - & - \\
        \hline
        q14 & $19.62$ & $25.21$ & $0.17$ & $3.33$  & $3.58$ & - & - \\
        \hline
  \end{tabular}
  \caption{\label{supp: tab__fit_parameter__chi} 
  \textbf{Device parameters obtained from the extended circuit model fit to the measured spectra (c.f. \figref{supp:fig__all_qubit_spectra}}).}
\end{table*}

\begin{center}
  \begin{figure}
    \includegraphics[width=1.0\textwidth]{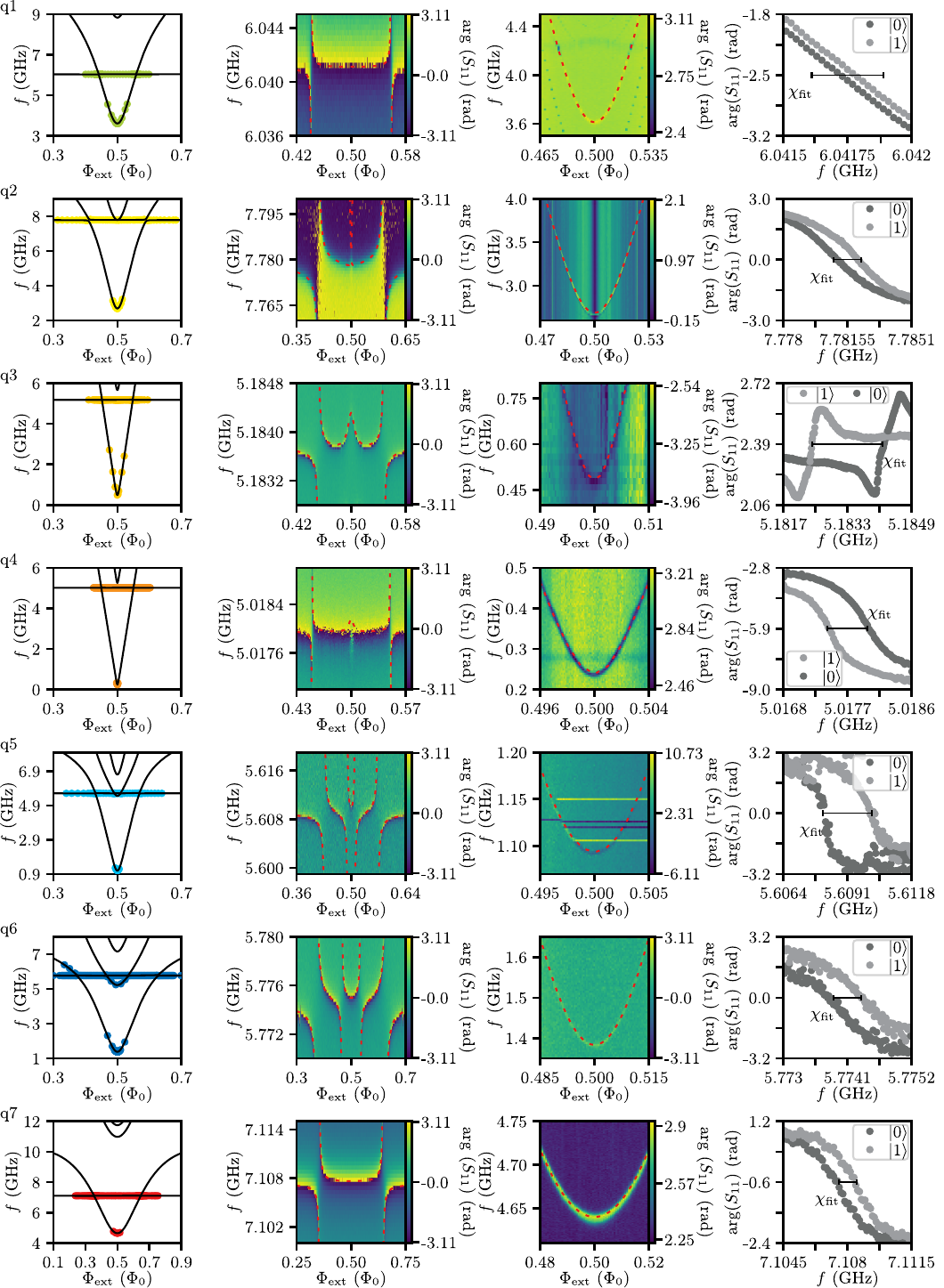}  
  \end{figure}
\end{center}
\begin{center}
  \begin{figure}
    \includegraphics[width=1.0\textwidth]{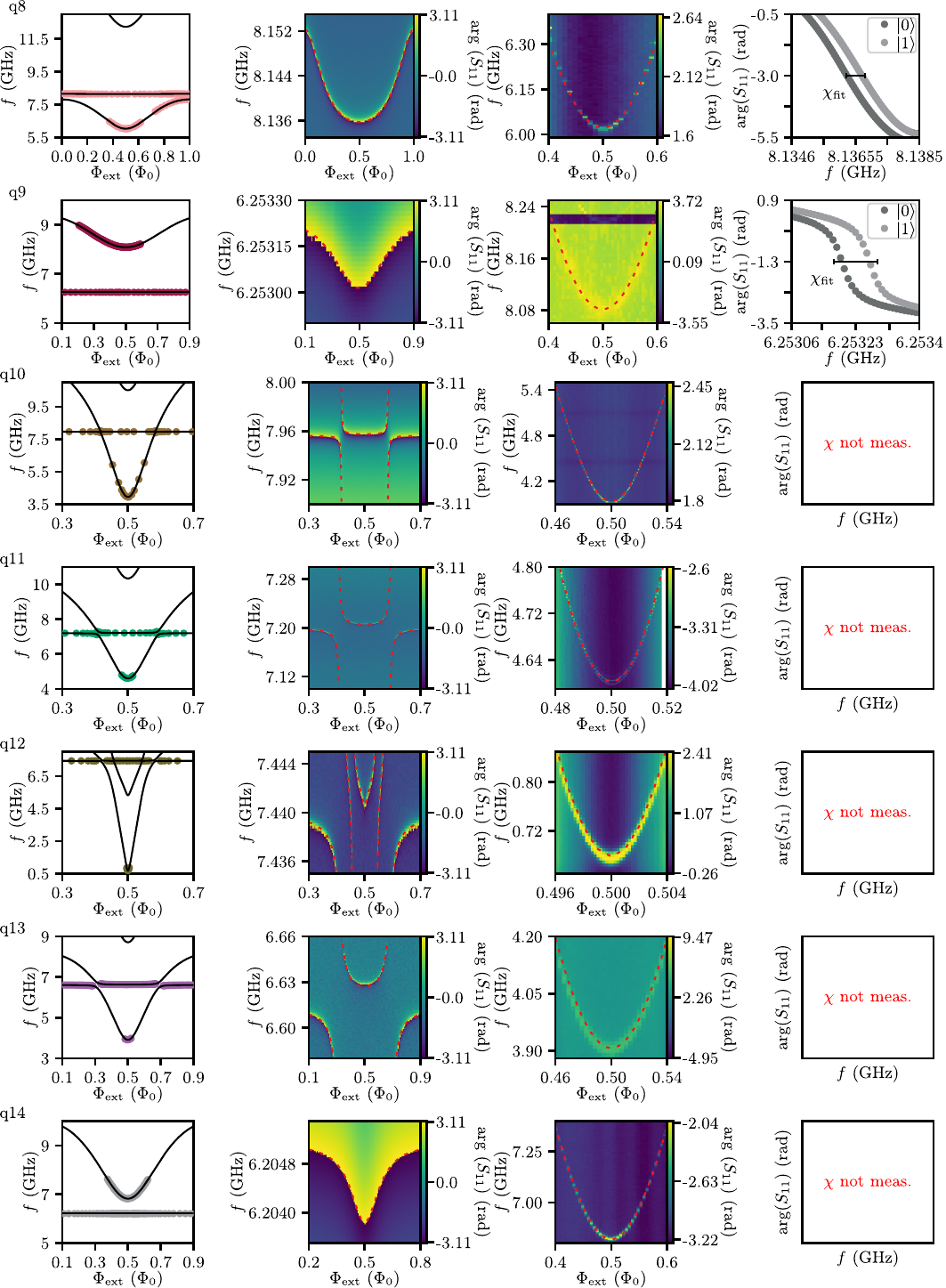}
  \end{figure}
\end{center}
\begin{center}
  \begin{figure}
    \caption{\label{supp:fig__all_qubit_spectra} 
    \textbf{Qubit spectra and dispersive shifts $\chi$ at $\Phi_\text{ext}=0.5\Phi_0$.} In the first column plots the data points are extracted from single-tone and two-tone spectroscopy of the resonator and qubit, shown in column 2 and 3, respectively. The black lines show fits to the circuit model.
    The measured dispersive shifts plotted in the 4$^\mathrm{th}$ column for devices q1 to q9 are extracted from pulsed single shot readout. The response of the resonator for the qubit in the ground (excited) state is shown in dark (light) grey markers. The $\chi$ value calculated from the circuit model is shown as a horizontal black line. The measured and calculated $\chi$ values are tabulated in Table \ref{supp: tab__fit_parameter__chi}.} 
  \end{figure}
\end{center}

\newpage

\subsection{\label{sec:electromagnetic_simulations} Electrostatic finite-element simulations}

\begin{figure*}[!htb]
    \includegraphics[width=1.0\textwidth]{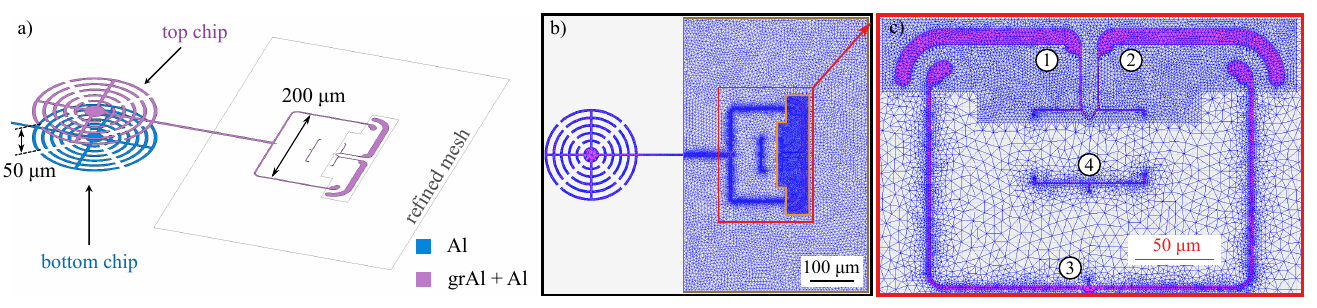}
    \caption{\label{fig:Ansys_sim} \textbf{Electrostatic finite element simulations with ANSYS Maxwell for the capacitance matrix $\mathbf{C}$.}   \textbf{a)} Layout of the flip-chip assembly. The qubit-readout circuit is flipped above the bottom chip to form the parallel plate capacitor $C_{\text{c}}$ (c.f. \figref{fig:GFQ} in the main text) with a distance of approx. 50 $\upmu$m between the chips. The readout line on the bottom chip is itself coupled capacitively to a coaxial cable (not shown) through which the system is measured in reflection.
    \textbf{b)} Mesh details of the qubit-readout system. The capacitive parts are colored in magenta. The mesh tetrahedra are superposed in blue. Refined mesh regions are used for the sensitive regions inside the brown and orange boxes. \textbf{b)} Zoom in on the qubit loop and capacitive structures of the qubit-readout system. The inductive parts and the junction are omitted in the simulation, leaving behind disconnected areas defining the circuit islands labeled by the node indices \textbf{1} to \textbf{4}. The mesh is generated automatically and is not perfectly symmetric with respect to islands \textbf{1} and \textbf{2}.}   
 \end{figure*}

To obtain the capacitance matrix $\mathbf{C}$, electrostatic simulations are conducted, for which the electrostatic finite element solver Ansys Maxwell is used. The simulations are performed with a detailed 3D model of the copper sample box with refined mesh regions used for finer structures, as can be seen in \figref{fig:Ansys_sim}. The mesh is automatically generated and therefore not perfectly symmetric with respect to the islands \textbf{1} and \textbf{2} (see \figref{fig:Ansys_sim}.b) leading to convergence errors of 10\,aF for the simulated capacitances, which are listed in Table \ref{tab:ANSYS_parasitic_capacitances} for all simulated devices. 

 \begin{table*}[!htb]\centering
    \begin{tabular}{|c||c|c|c|c|c|c|c|c|c|c|}
        \hline
        Qubit & $C_{11}$\,[fF] & $C_{22}$\,[fF] & $C_{33}$\,[fF] & $C_{44}$\,[fF] &  $C_{12,21} = C_{\text{sh}}$ \,[fF]  & $C_{24,42}$\,[fF]  & $C_{14,41}$\,[fF]  &  $C_{23,32} = C_{\text{r}}$\,[fF]   &$C_{13,31} = C_{\text{r}}$\,[fF]   & $C_{34,43}$\,[fF] \\
        \hline
        \hline
         q1 & $33.42$ & $33.28$ & $99.36$ & $2.92$ & $11.77$ & $0.29$ & $0.29$ & $8.63$ & $8.67$ & $0.84$ \\
        \hline
        q2 & $32.45$ &$32.39$ &  $76.93$ & $3.03$ & $3.22$ & $0.36$ & $0.32$ & $6.30$ & $6.34$ & $0.72$ \\
        \hline
        q3  & $16.47$ & $16.51$ & $68.69$ & $4.78$ & $3.38$ & $0.61$ & $0.60$ & $5.60$ & $5.59$ & $1.89$ \\
        \hline
        q4 & $16.44$ & $16.48$ & $68.89$ & $4.75$  & $3.40$ & $0.49$ & $0.48$ & $5.62$ & $5.61$ & $2.12$ \\
        \hline
        q5 & $33.84$ & $33.76$ & $74.27$  & $4.94$ & $4.20$ & $0.58$ & $0.57$ & $6.73$ & $6.76$ & $2.01$ \\
        \hline
        q6 & $33.59$ & $33.58$ & $76.00$ & $5.41$ & $4.29$ & $0.48$ & $0.48$ & $6.77$ & $6.79$ & $2.56$ \\
        \hline
        q7 & $16.07$ & $16.19$ & $71.19$ & $4.19$ & $3.27$ & $0.35$ & $0.34$ & $5.65$ & $5.61$ & $2.07$ \\
        \hline
        q8 & $53.37$ & $53.39$ & $126.97$ & $14.17$ & $5.67$ & $2.15$ & $2.13$ & $28.42$ & $28.44$ & $4.86$ \\
        \hline
        q9 & $17.30$ & $17.23$ & $73.67$ & $5.39$ & $3.72$ & $0.56$ & $0.57$ & $6.03$ & $6.04$ & $2.34$ \\
        \hline
        \hline
        q10 & $19.51$ & $19.52$ & $116.36$ & $8.95$ & $7.35$ & $0.56$ & $0.57$ & $5.35$ & $5.34$ & $4.03$ \\
        \hline
        q11 & $33.06$ & $33.22$ & $99.98$ & $2.90$ & $11.64$ & $0.27$ & $0.28$ & $8.71$ & $8.65$ & $0.83$ \\
        \hline
        q12 & $33.06$ & $33.10$ & $92.67$ & $2.75$ & $11.51$ & $0.29$ & $0.30$ & $7.90$ & $7.88$ & $0.77$ \\
        \hline
        q13 & $33.59$ & $33.60$ & $73.62$ & $4.33$ & $4.01$ & $0.61$ & $0.60$ & $6.75$ & $6.74$ & $1.61$ \\
        \hline
        q14 & $17.30$ & $17.23$ & $73.67$ & $5.39$ & $3.72$ & $0.56$ & $0.57$ & $6.03$ & $6.04$ & $2.34$ \\
        \hline
  \end{tabular}
  \caption{\textbf{Simulated capacitance values}. Capacitances depicted in  \figref{fig:full_circuit} simulated with the 3D electrostatic finite element solver Ansys Maxwell for different qubit designs with a convergence accuracy of 10\,aF. Differences in the qubit capacitances are due to different sizes of the circuit islands for different designs.
  \label{tab:ANSYS_parasitic_capacitances}}
\end{table*}


\subsection{\label{sec:coherence_times} Coherence times}
A summary of the measured coherence times for 11 GFQ devices is shown in Table \ref{tab:coherence}. We measure the free energy relaxation time $T_1$, Ramsey decoherence time $T_2^{*}$ and echo decoherence time $T_2^{\mathrm{echo}}$. Depending on the qubit, coherence times range from 1 - 10 $\upmu \mathrm{s}$ and for each qubit we observe fluctuations in time on the order of microseconds.

The most dominant loss channel for the generalized flux qubits presented in this study is inductive loss, which can be estimated by evaluating Fermi's Golden Rule via
\begin{equation}
    \frac{1}{T_1} = \frac{8\pi^3E_L}{hQ_{\text{ind}}} |\braket{0|\hat{\varphi}|1}|^2 \left( 1 + \coth{\frac{hf_{\text{q}}}{2k_{\text{B}}T}}\right),
\end{equation}
where $E_L$ is the inductive energy, $Q_{\text{ind}}$ the inductive quality factor, $\hat{\varphi}$ the flux operator in units of $\Phi_0$, and $f_{\text{q}}$ the qubit frequency. The measured energy relaxation times correspond to inductive quality factors of the order of $10^5$ to $10^6$ which is consistent with the expected inductive loss of granular aluminum~\cite{Gru__Loss_Mech_grAl__2018}. The quality factors calculated from the measured energy relaxation times are tabulated in the fifth column in Table \ref{tab:coherence}.

 \begin{table*}[!htb]\centering
    
    \begin{tabular}{|c||c|c|c|c|}
        \hline
        device & T$_1$ ($\upmu \mathrm{s}$) & T$_2^{*}$ ($\upmu \mathrm{s}$) & T$_2^{\mathrm{echo}}$ ($\upmu \mathrm{s}$) & $Q_{\text{ind}} (\times 10^6)$  \\
        \hline
        \hline
         q1 & $7.6$ / $9.8$ / $4.7$ / $3.2 \pm 0.3$ & $5.7$ / $2.1$ / $-$ &  $10.8$ / $12.6$ / $3.0$ &  0.9 / 1.16 / 0.56 / 0.38 \\
        \hline
         q2 & $10.3$  & $2.0$ & $-$ &  1.57 \\
        \hline
         q3 & $2.4$  & $-$ & $-$ &  1.31 \\
        \hline
         q4 & $1.3$  & $2.1$ & $-$ & 1.27 \\
        \hline
         q5 & $4.3$  & $2.4$ & $-$ &  0.90 \\
        \hline
         q7 & $8.0 \pm 2.4$  & $6.0$ & $7.7$ &  0.61 \\
        \hline
         q8 & $1.4$  & $2.3$ & $-$ &  0.07 \\
        \hline
         q10 & $4.1$ & $-$ & $-$ &  0.56  \\
        \hline
         q11 & $10.5$ / $9.9$ / $6.7$ / $6.2 \pm 0.4$ & $5.6$ / $2.6$ / $-$ & $-$ / $4.8$ / $10.2$ &  1.03 / 0.97  / 0.66 / 0.61 \\
        \hline
         q13 & $4.4$  & $1.4$ & $-$ &  0.28 \\
        \hline
         q15 & $15.3 \pm 2.1$  & $9.1$ & $-$ & $-$\\
        \hline

  \end{tabular}
  \caption{\textbf{Measured coherence times at the half-flux sweet spot for different samples.} For measurements with a statistically relevant amount of repetitions we show the mean value and standard deviation. Values separated by forward slashs are taken from different cooldowns, several months apart and measured in different sample holders. In between cooldowns the samples have been stored in ambient conditions.
  \label{tab:coherence}}
\end{table*}

\subsection{\label{sec:pulse} Pulse calibration}

\begin{figure*}[!htb]
    \includegraphics[width=0.714\textwidth]{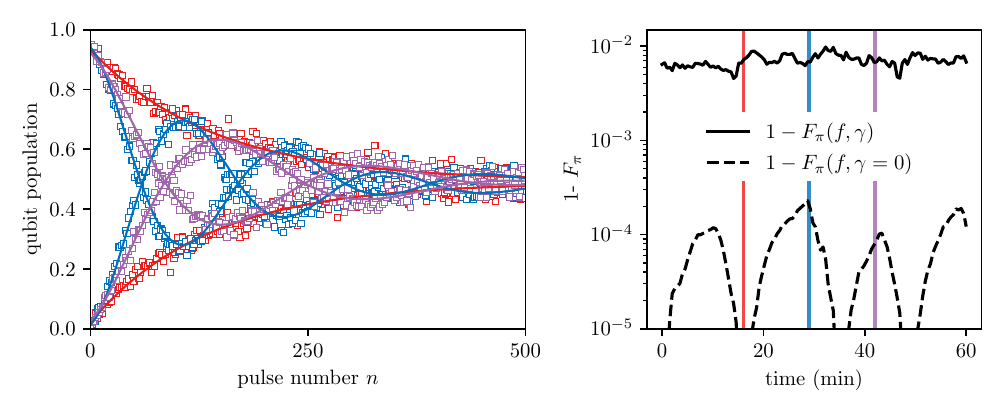}
    \caption{\label{fig:pi-pulse} \textbf{$\pi$-pulse calibration.  a)} The qubit population is plotted versus a subsequently played number of $\pi$-pulses for three individual experiments conducted 13 min apart. Oscillations in this measurement indicate the imperfection of the $\pi$-pulse over- or underscoring a perfect bit-flip that sum up to an inversion of the qubit population after half a period of the oscillation. The oscillations are dampened by energy relaxation.
    \textbf{b)} The experiment in \textbf{a} is conducted contiguously over the course of one hour and we plot $F_{\pi}(f, \gamma)$ and $F_{\pi}(f, \gamma = 0)$ in continuous and dashed lines, respectively. $F_{\pi}(f, \gamma = 0)$ drifts continuously with a period of tens of minutes.
    }
\end{figure*}

We fine-tune the $\pi$-pulse amplitude by minimizing the beating pattern in the qubit population measured after a sequence of $n$ successive pulses. Three typical measurements are shown in \figref{fig:pi-pulse}a. The qubit population vs. pulse number is given by
\begin{equation}
    P(n) = a \left( \frac{1}{2} - \frac{1}{2}\cos(\pi n + 2\pi f n)     \right) \exp(-\gamma n) \, + o,
    \label{eq:pi-pulse-fidelity}
\end{equation} 
where $f$ is the frequency of the beating and $\gamma$ accounts for energy decay. The amplitude $a$ and offset $o$ account for measurement errors.  Using the parameters extracted from a fit of the data using \eqref{eq:pi-pulse-fidelity}, we define the $\pi$- pulse fidelity as 
\begin{equation*}
    F_{\pi}(f, \gamma) = (P(1) - P(0))/a = \left( \frac{1}{2} - \frac{1}{2}\cos(\pi + 2\pi f)    \right) \exp(-\gamma).
\end{equation*}
In \figref{fig:pi-pulse}.b we plot $F_{\pi}(f, \gamma)$ and $F_{\pi}(f, \gamma = 0)$ for repeated measurements over one hour. While the pulse calibration error $F_{\pi}(f, \gamma = 0)$ drifts in time, the dominant error source is energy decay.

\begin{figure*}[!htb]
    \includegraphics[width=0.8\textwidth]{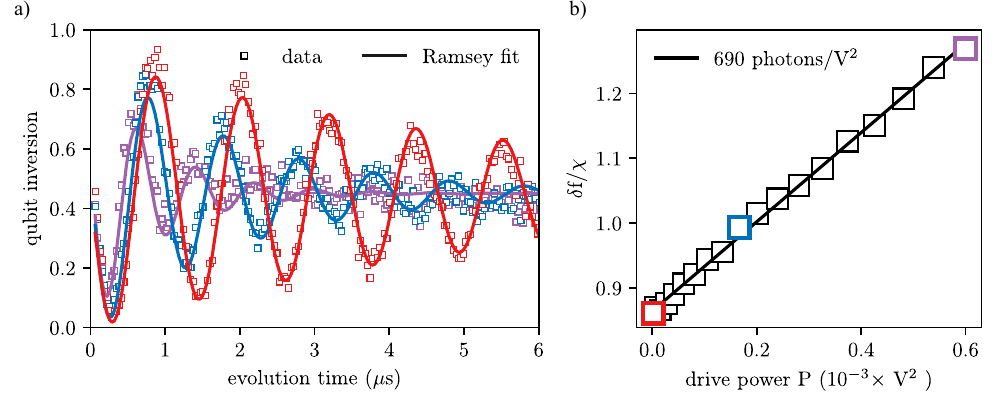}
    \caption{\label{fig:photon} \textbf{Photon number calibration.} \textbf{a)} Ramsey fringes are recorded while simultaneously populating the readout resonator using different drive powers. The fringes are fitted to a sinusoidal function (continuous lines) to obtain the qubit frequency detuning $\delta f$ with respect to the qubit pulse. The red markers correspond to no drive and the powers corresponding to the blue and violet points are the same as in panel b.
    \textbf{b)} The linear fit of $\delta f$ vs. different readout powers (continuous line) gives the photon number calibration used in the main text.  
    }
\end{figure*}

The photon number for the figures in the main text is calibrated using Ramsey interferometry while populating the readout resonator as depicted in \figref{fig:photon}.a. We use the linear AC-Stark shift in the dispersive regime
\begin{equation}
    \overline{n} = \frac{\delta f}{\chi} \propto P,
\end{equation}
where $\delta f$ is the frequency shift of the qubit and $\chi$ is the dispersive shift of the resonator, to relate the circulating photon number $\overline{n}$ to the power of the readout drive $P$ by fitting the Ramsey fringes for different $P$. We apply a linear fit as it is depicted in \figref{fig:photon}.b.

\end{document}